\documentclass[preprint,showpacs,preprintnumbers,amsmath,amssymb]{revtex4}
\usepackage{graphicx}
\usepackage{dcolumn}
\usepackage{bm}

\newcommand{\be}{\begin{equation}}
\newcommand{\en}{\end{equation}}
\newcommand{\bea}{\begin{eqnarray}}
\newcommand{\ena}{\end{eqnarray}}

\begin{document}

\title{Intermediate inflation  in Gauss-Bonnet braneworld}

\author{Ram\'{o}n Herrera\footnote{E-mail address: ramon.herrera@ucv.cl}}
\affiliation{Instituto de F\'{\i}sica, Pontificia Universidad
Cat\'{o}lica de Valpara\'{\i}so, Avenida Brasil 2950, Casilla
4059, Valpara\'{\i}so, Chile.}
\author{Nelson Videla\footnote{E-mail address: nelson.videla@ucv.cl}}
\affiliation{Instituto de F\'{\i}sica, Pontificia Universidad
Cat\'{o}lica de Valpara\'{\i}so, Avenida Brasil 2950, Casilla
4059, Valpara\'{\i}so, Chile.}
\date{\today}

\begin{abstract}

 In this article we study  an intermediate
inflationary universe models using the Gauss-Bonnet brane. General
conditions required for these models
 to be realizable are derived and
 discussed. We use
recent astronomical observations  to constraint  the parameters
appearing in the model.

\end{abstract}

\maketitle

\section{\label{sec:level1} Introduction}
It is well known that one of the most exciting ideas of
contemporary physics is to explain the origin of the observed
structures in our universe. It is believed that Inflation
\cite{R1} can provide an elegant mechanism to explain the
large-scale structure, as a result of quantum fluctuations in the
early expanding universe, predicting that small density
perturbations are likely to be generated in the very early
universe with a nearly scale-free spectrum \cite{R2}. This
prediction has been supported by early observational data,
specifically in the detection of temperature fluctuations in the
cosmic microwave background (CMB) by the COBE satellite \cite{R4}.
The scheme of inflation \cite{IC} (see \cite{libro} for a review)
is based on the idea that at early times there was a phase in
which the universe evolved through accelerated expansion in a
short period of time at high energy scales. During this phase, the
universe was dominated by a potential $V(\phi)$ of a scalar field
$\phi$ (inflaton).


In the context of inflation we have the particular scenario of
''intermediate inflation'', in which  the scale factor evolves as
$a(t)=\exp(A t^f)$. Therefore, the expansion of the universe is
slower than standard de Sitter inflation ($a(t)=\exp(H t)$), but
faster than power law inflation ($a(t)= t^p; p>1$). The
intermediate inflationary model was introduced as an exact
solution for a particular scalar field potential of the type
$V(\phi)\propto \phi^{-4(f^{-1}-1)}$, where $f$ is a free
parameter\cite{Barrow1}. Recently, a  tachyon field in
intermediate inflation was considered in \cite{yo3}, and a
 warm-intermediate inflationary universe model was
studied en Ref.\cite{yo4} (see also Ref.\cite{yo5}).

The  motivation to study intermediate inflationary model becomes
from string/M theory (for a review see Refs.\cite{1}). This theory
suggests that in order to have a ghost-free action high order
curvature invariant corrections to the Einstein-Hilbert action
must be proportional to the Gauss-Bonnet (GB) term\cite{17}. GB
terms arise naturally as the leading order of the  expansion to
the low-energy string effective action, where  is the inverse
string tension\cite{18}. This kind of theory has been applied to
possible resolution of the initial singularity problem\cite{19},
to the study of Black- Hole solutions\cite{20}, accelerated
cosmological solutions\cite{21}, among others (see
Refs.\cite{Meng:2003pn,Lidsey:2003sj, Calcagni:2004as, Kim:2004gs,
Kim:2004jc, Kim:2004hu, Murray:2006fw}). In particular, very
recently, it has been found that for a dark energy model the GB
interaction in four dimensions with a dynamical dilatonic scalar
field coupling leads to a solution of the form $a = \exp At^f$
\cite{22}, where the universe starts evolving with a decelerated
exponential expansion. Here, the constant $A$ becomes given by $A
= \frac{2}{\kappa n}$ and $f = \frac{1}{2}$ , with $\kappa^2 = 8
\pi G$ and $n$ is a constant. Also, much attention has been
focused on the Randall Sundrum (RS) scenario, where our observable
four-dimensional universe is modelled as a domain wall embedded in
a higher-dimensional bulk space \cite{2}. These kind of models can
be obtained from superstring theory~\cite{witten, witten2}. For a
comprehensible review on RS cosmology, see
Refs.~\cite{lecturer,lecturer2,lecturer3}. In this way, the idea
that inflation , or specifically, intermediate inflation, comes
from an effective theory at low dimension of a more fundamental
string theory is in itself very appealing. Thus, in brane universe
models the effective theories that emerge from string/M theory
lead to a Friedmann Equation which is proportional to the square
energy density.




 When the five dimensional Einstein-GB equations are projected on
to the brane, a complicated Hubble equation  is obtained \cite{Ch,
Davis:2002gn, Gravanis:2002wy}. Interestingly enough, this
modified Friedmann equation reduces to a very simple equation
$H^2\propto \rho^q$ with $q = 1, 2, 2/3$ corresponding to General
Relativity (GR), RS and GB regimes, respectively. This situation
motivated the "patch cosmology" as a useful approach to study
braneworld scenarios \cite{Calcagni:2004bh}. This scheme makes use
of a nonstandard Friedmann equation of the form $H^2= \beta_q^2
\rho^q$. Despite all the shortcomings of this approximate
treatment of extra-dimensional physics, it gives several important
first-impact information. Recently, a closed inflationary universe
in patch cosmology was considered in \cite{yo1}, and a tachyonic
universes in patch cosmologies with $\Omega>1$ was studied in
Ref.\cite{yo2}.

The purpose of the present work is to study intermediate
inflationary universe models, where the matter content is confined
to a four dimensional brane which is embedded in a five
dimensional bulk where a GB contribution is considered. We study
these models using the approach of patch cosmology.  On the other
hand, a comprehensive study in the present work reveals that,
intermediate inflation provides  the possibility of density
perturbation and gravitational wave spectra which differ from the
usual inflationary prediction of a nearly flat spectrum with
negligible gravitational waves.  Furthermore, in the present model
the tensor-to-scalar ratio $r$ is scale-dependent, and we have
shown that a good fit to the WMAP5 observations.

The outline of the Letter is as follows. The next section we
briefly review the cosmological equations in the GB brane world
and present the patch cosmological equations for this model. In
Sect. III presents a short review of the intermediate inflation in
GB brane.  In Sect. IV the cosmological perturbations are
investigated. Finally, in Sect. V we summarize our finding.

\section{\label{Sec1} Cosmological Equations in Gauss-Bonnet brane}

We start with the five-dimensional bulk action for the GB
braneworld:

\begin{eqnarray}
S &=&\frac{1}{2\kappa
_{5}^{2}}\int_{bulk}d^{5}x\sqrt{-g_{5}}\left\{ R-2\Lambda
_{5}+\alpha \left( R^{\mu \nu \lambda \rho }R_{^{\mu \nu \lambda
\rho }}-4R^{\mu \nu }R_{\nu \mu }+R^{2}\right) \right\} \nonumber \\
&&+\int_{brane}d^{4}x\sqrt{-g_{4}}\left(
\mathcal{L}_{matter}-\sigma \right),\label{action1}
\end{eqnarray}
where $\Lambda _{5}=-3\mu ^{2}\left( 2-4\alpha \mu ^{2}\right) $
is the cosmological constant in five dimensions, with the
$AdS_{5}$ energy scale $\mu$, $\alpha$ is the GB coupling
constant, $\kappa _{5}=8\pi/m_{5}$ is the five dimensional
gravitational coupling constant and $\sigma$ is the brane tension.
 $\mathcal{L}_{matter}$ is the matter lagrangian for the
inflaton field on the brane. We will consider the case that a
perfect fluid matter source with density $\rho$ is confined to the
brane.

A Friedmann-Robertson-Walker (FRW) brane in an AdS$_5$ bulk is a
solution to the field and junction equations (see Refs.\cite{Ch,
Davis:2002gn, Gravanis:2002wy}). The modified Friedmann on the
brane can be written as
 \begin{equation}
H^2 = {1\over
4\alpha}\left[(1-4\alpha\mu^2)\cosh\left({2\chi\over3}
\right)-1\right]\,,\label{q22}
 \end{equation}
\begin{equation}
\label{var1} \kappa_5^2(\rho+\sigma) =
\left[{{2(1-4\alpha\mu^2)^3} \over {\alpha} }\right]^{1/2}
\sinh\chi\,,
\end{equation}
where $\chi$ represents a dimensionless measure of the energy
density $\rho$.
In this work we will assume that the matter fields are restricted
to a lower dimensional hypersurface (brane) and that gravity
exists throughout the space-time (brane and bulk) as a dynamical
theory of geometry. Also, for 4D homogeneous and isotropic
Friedmann cosmology, an extended version of Birkhoff's theorem
tells us that if the bulk space-time is AdS, it implies that  the
effect of the Weyl tensor (known as dark radiation) does not
appear in the modified Friedmann equation. On the other hand, the
brane Friedmann equation for the general,  where the bulk
spacetime may be interpreted as a charged black hole was studied
in Refs.\cite{Od,Od2,Od3}.

The modified Friedmann equation~(\ref{q22}), together with
Eq.~(\ref{var1}), shows that there is a  characteristic
Gauss-Bonnet energy scale\cite{pp}
 \be \label{gbscale}
m_{GB}= \left[{{2(1-4\alpha\mu^2)^3} \over {\alpha} \kappa_5^4
}\right]^{1/8}\,,
 \en
such that the GB high energy regime ($\chi\gg1$) occurs if
$\rho+\sigma \gg m_{GB}^4$.  Expanding Eq.~(\ref{q22}) in $\chi$
and using (\ref{var1}), we find in the full theory three regimes
for the dynamical history of the brane universe \cite{Ch,
Davis:2002gn, Gravanis:2002wy}:

 \be
\rho\gg m_{GB}^4~ \Rightarrow ~ H^2\approx \left[ {\kappa_5^2
\over 16\alpha}\, \rho
\right]^{2/3}\,\;\;\;\;\;\;\;(GB),\label{gbl}
 \en
 \be
 m_{GB} \gg
\rho\gg\sigma  \Rightarrow ~ H^2\approx {\kappa_4^2 \over
6\sigma}\, \rho^{2}\,\;\;\;\;\;\;\;\;\;\;(RS),\label{rsl}
 \en
 \be
\rho\ll\sigma~ \Rightarrow ~ H^2\approx {\kappa_4^2 \over 3}\,
\rho\,\;\;\;\;\;\;\;\;\;\;\;\;\;\;\;\;\;\;\;\;\;\;\;(GR).
\label{ehl}
 \en
Clearly Eqs. (\ref{gbl}), (\ref{rsl}) and (\ref{ehl}) are much
simpler than the full Eq (\ref{q22}) and in a practical case one
of the three energy regimes will be assumed. Therefore, patch
cosmology can be useful to describe the universe in a region of
time and energy in which \cite{Calcagni:2004bh}
\begin{equation}
H^{2}=\beta_{q}^2\rho ^{q}, \label{dda}
\end{equation}
where $H=\dot{a}/a$ is the Hubble parameter  and $q$ is a patch
parameter that describes a particular cosmological model under
consideration. The choice $q=1$ corresponds to the standard
General Relativity with $\beta^2_1=8\pi/3m_{p}^2$, where $m_{p}$
is the four dimensional Planck mass. If we take $q=2$, we obtain
the high energy limit of the brane world cosmology, in which
$\beta^2_2=4\pi/3\sigma m_p^2 $. Finally, for $q=2/3$, we have the
GB brane world cosmology, with $\beta^2_{2/3}=G_{5}/16\zeta$,
where $G_5$ is the $5D$ gravitational coupling constant and
$\zeta=1/8g_s$ is the GB coupling ($g_s$ is the string energy
scale). The parameter $q$, which describes the effective degrees
of freedom from gravity, can take a value in a non-standard set
because of the introduction of non-perturbative stringy effects.
Here, we mentioned some possibilities, for instance, in
Ref.\cite{Kim:2004hu} it was found that an appropriate region to a
patch parameter $q$ is given by $1/2 = q < \infty$. On the other
hand, from Cardassian cosmology it is possible to obtain a
Friedmann equation similar  (\ref{dda}) as a consequence of
embedding our observable universe as a 3+1 dimensional brane in
extra dimensions. In fact, in Ref.\cite{Chung:1999zs} a modified
FRW equation  was obtained in our observable brane with $H^2
\propto \rho^n$ for any $n$.


On the other hand, we neglect any contribution from both the Weyl
tensor and the brane-bulk exchange, assuming there is some
confinement mechanism for a perfect fluid. Thus, the energy
conservation equation on the brane follows directly from the
Gauss-Codazzi equations. For a perfect fluid matter source it is
reduced to the familiar form, $ \dot{\rho}+3H\left( \rho +P\right)
=0,$  where $\rho$ and $P$ represent the energy and pressure
densities, respectively. The dot denotes derivative with respect
to the cosmological time $t$.

 We consider that the matter content of the universe is a homogeneous
inflaton field $\phi(t)$ with potential $V(\phi)$. Then the energy
density and pressure are given by $
\rho=\frac{\dot{\phi}^2}{2}+V(\phi)$ and
$P=\frac{\dot{\phi}^2}{2}-V(\phi)$, respectively. In this way, the
equation of motion of the rolling scalar field becomes
\begin{equation}
\displaystyle \ddot{\phi}+3\,H\,\dot{\phi}+V'(\phi )=0\,\,,
\label{ecphi}
\end{equation}
here,  for convenience we will use  units in which $c=\hbar=1$ and
$V'(\phi )=\partial\,V(\phi)/\partial\phi$.

\section{Intermediate inflation in Gauss Bonnet Brane}
In this section  exact solutions can be found for intermediate
inflationary universes models where the scale factor,\ $a(t)$,
expands as follows

\begin{equation}
a(t)=\exp(At^{f}).\label{at}
\end{equation}
Here $f$\ \ is a constant parameter with range $0<f<1$\ , and $A$\
is a positive constant.

From equations (\ref{dda}), (\ref{ecphi}), and using (\ref{at}),
we obtain
\begin{equation}
\dot{\phi}^{2}=-\frac{2\dot{H}H^{2\alpha_{0}}}{3q\beta_{q}^{2(\alpha_{0}+1)}%
}=\frac{2(Af)^{2\alpha_{0}+1}(1-f)}{3q\beta_{q}^{2(\alpha_{0}+1)}}\;\;
t^{2\alpha_{1}},\label{dp}
\end{equation}
and the effective potential as a function of the cosmological
times becomes

\begin{equation}
V(t)=\left[  \frac{\left(  Af\right)
^{2}t^{2(f-1)}}{\beta_{q}^{2}}\right] ^{\frac{1}{q}}-\frac{\left(
Af\right)  ^{2\alpha_{0}+1}(1-f)}{3q\beta
_{q}^{2(\alpha_{0}+1)}}t^{2\alpha_{1}},\label{Vt}
\end{equation}
where
$$
\alpha_{0}    =\frac{1-q}{q},\;\;\;\; \mbox {and}\,\;\;\;
\alpha_{1} =f(\alpha_{0}+\frac{1}{2})-(\alpha_{0}+1),
$$
are constant  parameters, respectively.

The solution for the scalar field $\phi(t)$  can be found from
Eq.(\ref{dp})

\begin{equation}
(\phi-\phi_{0})=A^{\alpha_{0}+\frac{1}{2}}\alpha_{2}t^{\alpha_{1}+1},\label{fpt}
\end{equation}
where $\phi(t=0)=\phi_{0}$. Here, the parameter $\alpha_2$ is
defined by
$$
\alpha_{2}=\left[
\frac{2f^{2\alpha_{0}+1}(1-f)}{3q\beta_{q}^{2(\alpha
_{0}+1)}(\alpha_{1}+1)^{2}}\right]  ^{\frac{1}{2}}.
$$

An exact solution  of Eqs. (\ref{Vt}) and (\ref{fpt}) of the form
of Eq.(\ref{at}) exists with

\begin{equation}
V(\phi)=A^{\frac{2\alpha_{3}}{q}}\alpha_{4}(\phi-\phi_{0})^{\frac
{2(f-1)}{q(\alpha_{1}+1)}}-\frac{A^{\frac{2(\alpha_{0}+\frac{1}{2})}%
{\alpha_{1}+1}}}{2}\alpha_{5}(\phi-\phi_{0})^{\frac{2\alpha_{1}}{\alpha_{1}%
+1}},\label{Vf}
\end{equation}
where
$$
\alpha_{3}
=1+\frac{(\alpha_{0}+\frac{1}{2})(1-f)}{(\alpha_{1}+1)},\;\;
\alpha_{4}   =\left[  \frac{f^{2}\alpha_{2}^{\frac{2(1-f)}{\alpha_{1}+1}}%
}{\beta_{q}^{2}}\right]  ^{\frac{1}{q}},\,\,\mbox{and}\;\;\;
\alpha_{5} =\alpha_{2}^{\frac{2}{\alpha_{1}+1}}(\alpha_{1}+1)^{2}.
$$

The Hubble parameter as a function of the inflaton field \ $\phi$\  becomes%
\begin{equation}
H(\phi)=A^{\alpha_{3}}f\alpha_{2}^{\frac{1-f}{\alpha_{1}+1}}(\phi-\phi
_{0})^{\frac{f-1}{\alpha_{1}+1}}.\label{H}
\end{equation}

Assuming the set of slow-roll conditions,
$\frac{\dot{\phi}^{2}}{2}\ll V(\phi)$
and\ $\ddot{\phi}\ll 3H\dot{\phi}$, the potential given by Eq.(\ref{Vf}) reduces to%

\begin{equation}
V(\phi)=A^{\frac{2\alpha_{3}}{q}}\alpha_{4}(\phi-\phi_{0})^{\frac
{2(f-1)}{q(\alpha_{1}+1)}}.\label{pot}
\end{equation}
Here, the first term of the effective potential given by  Eq.
(\ref{Vf}) dominates  at large values of \ $(\phi-\phi_{0})$. Note
that, the solutions for $\phi(t)\ $ and $H(\phi)$, corresponding
to this potential are identical to those obtained when the exact
potential, Eq. (\ref{Vt}), is used.

We should note that in the GR regime, i.e., $q=1$ the scalar
potential becomes $V(\phi)\propto\phi^{-4(1-f)/f}$, in the RS
regime i.e., $q=2$, the potential is
$V(\phi)\propto\phi^{-2(1-f)}$, and finally in the GB regimen
$q=2/3$, $V(\phi)\propto\phi^{-3(1-f)/(f-1/2)}$. Without loss of
generality $\phi_0$ can be taken to be zero.  Note that the
potentials which are asymptotically of inverse power-law type are
commonly  used in quintessence models \cite{13}, but it also
establishes  viable inflationary solutions. These potentials also
arises from the scalar-tensor gravity theories\cite{14}.

Introducing the Hubble slow-roll parameters
$(\epsilon_{1},\eta_{\eta})$ and potential slow-roll parameters
$(\epsilon_{1}^{q},\eta_{n}^{q})$, see Ref.\cite{Kim:2004gs}, we
write

\begin{equation}
\epsilon_{1}=-\frac{\dot{H}}{H^{2}}\approx\epsilon_{1}^{q}=\frac
{qV^{\prime^{2}}}{6\beta_{q}^{2}V^{q+1}}=A^{2\alpha_{0}\alpha_{3}}\alpha
_{6}\phi^{-2\gamma},%
\end{equation}
and \
\begin{equation}
\eta_{n}    =-\frac{1}{H^{n}\dot{\phi}}\frac{d^{n+1}\phi}{dt^{n+1}}%
\approx\eta_{n}^{q},
\end{equation}
where
$$
\eta_{1}^{q}    =\frac{1}{3\beta_{q}^{2}}\left[  \frac{V^{\prime\prime}%
}{V^{q}}-\frac{qV^{\prime^{2}}}{2V^{q+1}}\right]  =A^{2\alpha_{0}\alpha_{3}%
}\alpha_{7}\phi^{-2\gamma},
$$
$$
 \eta_{2}^{q}  =\frac{-1}{(3\beta_q^2)^2}\left[\frac{V'V''}{V^{2q}}+\frac{(V'')^2}{V^{2q}}-\frac{5qV''(V')^2}{V^{2q+1}}
 +\frac{q(q+2)(V')^4}{2\,V^{2(q+1)}}\right]
$$
$$
=\left[\frac{-a^2\,B^{2(1-q)}\,\phi^{2a(1-q)-4}}{(3\beta_q^2)^2}\right]\,[1+a^2(1+q[q-8]/2)-\phi+a(\phi+5q-2)].
$$
Here, the parameters $\gamma$, $\alpha_6$, $\alpha_7$, $a$ and $B$
are
$$
\gamma   =\frac{\alpha_{0}(1-f)}{(\alpha_{1}+1)}+1,\;\;\;\;
\alpha_{6} =\frac{q\alpha_{4}^{1-q}}{6\beta_{q}^{2}}\left[ \frac
{2(1-f)}{q(\alpha_{1}+1)}\right]  ^{2},
$$
$$
\alpha_{7}   =\frac{2(1-f)\alpha_{4}^{1-q}}{q(\alpha_{1}+1)3\beta_{q}^{2}%
}\left[  \frac{(1-f)}{(\alpha_{1}+1)}\left(  \frac{q}{2}-1\right)
+1\right],
\;\;\;a=\frac{2(f-1)}{q(\alpha_1+1)},\;\;\mbox{and}\,\,\,B=\alpha_4\,A^{2\alpha_3/q},
$$
respectively.

Note that, the ratio between $\eta_{1}^{q}$\ and $\epsilon_{1}^{q}$\ becomes%
\begin{equation}
\frac{\eta_{1}^{q}\ }{\epsilon_{1}^{q}\ }=\frac{2}{q}+\frac{\alpha_{1}+1}%
{1-f}-1,
\end{equation}
and $\eta_{1}^{q}$ reaches unity before $\epsilon_{1}^{q}$ does.
Therefore,  we may establish that the end of inflation is governed
by the condition $\eta_{1}^{q}=1$\ in place of $\
\epsilon_{1}^{q}=1$. From this, we get for the scalar field $\phi$
at the end of inflation, becomes


\begin{equation}
\phi_{end}=\left(  A^{2\alpha_{0}\alpha_{3}}\alpha_{7}\right)
^{\frac {1}{2\gamma}}.\label{fe}
\end{equation}

On the other hand, the number of e-folds at the end of inflation
using Eqs. (\ref{dda}), (\ref{ecphi}) and (\ref{pot}) under the
set of slow-roll conditions, is given by
\begin{equation}
N=\int_{t_{\ast}}^{t_{end}}Hdt=3\beta_{q}^{2}\int_{\phi_{end}}^{\phi_{\ast}%
}\frac{V^{q}}{V^{\prime}}d\phi=A^{-2\alpha_{0}\alpha_{3}}\alpha_{8}\left\{
\phi_{end}^{2\gamma}-\phi_{\ast}^{^{2\gamma}}\right\}  ,\label{N}
\end{equation}
where
$$
\alpha_{8}=\frac{3\beta_{q}^{2}\alpha_{4}^{q-1}q(\alpha_{1}+1)}{4(1-f)\gamma}.%
$$
The subscripts $\ast$\ and $end$\ are used to denote the epoch
when the cosmological scales exit the horizon and the end of
inflation, respectively.

\section{\label{Sec3}  Perturbation spectral from intermediate inflation in patch cosmological models }
In this section we will study the scalar and tensor perturbations
for our model.  It has long been recognized that inflation gives
rise to a spectrum of scalar perturbations close to the
scale-invariant Harrison-Zel'dovich. For a scalar field the
amplitude of scalar perturbations generated during inflation for a
flat space is approximately \cite{Kim:2004gs}

\begin{equation}
\mathcal{P}_{\mathcal{R}}=\left(
\frac{H^{2}}{2\pi\dot{\phi}}\right)  _{k=k_{\ast}}^{2}=
A^{2\xi}\alpha_{9}\phi_{\ast}^{2\sigma},\label{Ps}
\end{equation}
where%
$$
\alpha_{9}
=\frac{f^{4}\alpha_{2}^{\frac{4(1-f)}{\alpha_{1}+1}}}{4\pi
^{2}\alpha_{5}},\,\,\;\; \;\;\xi   =2\alpha_{3}-\frac{\left(
\alpha_{0}+\frac{1}{2}\right) }{(\alpha
_{1}+1)},\,\,\;\;\mbox{and}\;\;\;\; \sigma =\frac{\left[
2(f-1)-\alpha_{1}\right] }{(\alpha_{1}+1)},
$$
respectively. Here we have used Eqs. (\ref{dp}),  and (\ref{H}).
The quantity   $k_{\ast}$, is refereed to $k=Ha$, the value when
the universe scales crosses the Hubble Horizon during inflation.

From Eqs. (\ref{fe}), (\ref{N}) and (\ref{Ps}), we obtained a
constraint for the parameter $A$ given by
\begin{equation}
A=\left\{  \frac{\mathcal{P}_{\mathcal{R}}}{\alpha_{9}}\left[  \frac{N}%
{\alpha_{8}}-\alpha_{7}\right]  ^{-\frac{\sigma}{\gamma}}\right\}
^{\frac {1}{2\left[
\xi+\alpha_{0}\alpha_{3}\frac{\sigma}{\gamma}\right]  }}.\label{A}
\end{equation}
 In this way, we can obtain the value of $A$ for a given values of
 $f$ and $\beta_q^2$ parameters when number of e-folds $N$, and the
 power spectrum of the curvature perturbations
 $\mathcal{P}_{\mathcal{R}}$ is given. Now we consider the special case in which
  $f=2/3$.  In this special case  we obtained that in GR
 ($q=1$) we get $A\simeq 0.0019m_p^{2/3}$. In RS ($q=2$) for the value of
 $\beta_{2}^2=10^{-13}m_p^{-6}$, we have $A\simeq 0.0014
 m_p^{2/3}$, and in the  GB regime ($q=2/3$)  for
 $\beta_{2/3}^2=10^{-3}m_p^{-2/3}$, we get $A\simeq 0.0046
 m_p^{2/3}$. Here we have taken $N=60$ and $\mathcal{P}_{\mathcal{R}}\simeq 2.4\times
 10^{-9}$.

Note that the general expression for the amplitude of scalar
perturbations in GB brane world is given by\cite{pp}
\begin{equation}
\mathcal{P}_{\mathcal{R}}=\left[\frac{\kappa_4^6\,V^3}{6\pi^2\,V'^2}\right]\,G_\beta^2(H/\mu)\;_{k=k_{\ast}},\label{P}
\end{equation}
where the term in square brackets is the standard scalar
perturbation, and the GB brane world correction is given by
$$
G_\beta^2(x)=\left[\frac{3\,(1+\beta)\,x^2}{2(3-\beta+2\beta\,x^2)\sqrt{1+x^2}+2(\beta-3)}\right]^3,
$$
where $x\equiv \,H\mu$ is a dimensionless measure of energy scale,
and $\beta=4\alpha\mu$. The RS amplification factor is recovered
when $\beta=0$\cite{Maartens:1999hf}.

We also consider the  q-spectral index $n_s^q$, which is related
to the power spectrum of density perturbations
$\mathcal{P}_{\mathcal{R}}$. For modes with a wavelength much
larger than the horizon ($k \ll a H$), where $k$ is the comoving
wave number. The scalar q-spectral index is given by $n_s^q=1+d
\ln \mathcal{P}_{\mathcal{R}}/d\ln k$, see
Ref.\cite{Murray:2006fw}, and in our case becomes

\begin{equation}
n_{s}^{q}=1-4\epsilon_{1}^{q}+2\eta_{1}^{q}=1-\frac{2A^{2\alpha_{0}\alpha_{3}%
}\alpha_{4}^{1-q}}{3\beta_{q}^{2}}\left[  \frac{2(1-f)}{q(\alpha_{1}%
+1)}\right]  \left[  \frac{(1-f)}{(\alpha_{1}+1)}\left(
3-\frac{2}{q}\right) -1\right]  \phi^{-2\gamma}\label{ns}.
\end{equation}

In order to confront these models with observations, we need to
consider the $q$-tensor-scalar ratio
$r_q=16\,A_{T,q}^2/A_{S,q}^2$, where the $q$-scalar amplitude is
normalized by $A_{S,q}^2=4\mathcal{P}_{\mathcal{R}}/25$. Here, the
tensor amplitude is given by
\begin{equation}
 A_{T,q}^2=A_{T,
GR}^2\,F_\beta^{2}(H/\mu),\label{At}
\end{equation}
 where  $A_{T,GR}^2$
is the standard amplitude in GR i.e.,
$A_{T,GR}=24\beta_1^2\,(H/2\pi)^2$, and the function $F_\beta$
contains the information about the GB term \cite{pp}
$$
F_\beta^{-2}=\sqrt{1+x^2}-\left(\frac{1-\beta}{1+\beta}\right)
\,x^2\,\sinh^{-1}\left(\frac{1}{x}\right)\;\;\;\;\;(x\equiv\frac{H}{\mu}).
$$

Following, Ref.\cite{Calcagni:2004bh}  we approximate the function
$F_\beta^2\approx F_q^2$, where for the GR regime
$F_{q=1}^2\approx F_\beta^2(H/\mu\ll 1)=1$, for  the RS regime
$F_{q=2}^2\approx F_{\beta=0}^2(H/\mu\gg 1)=3H/(2\mu)$, and
finally for the GB regime $F_{q=2/3}^2\approx F_\beta^2(H/\mu\gg
1)=(1+\beta)H/(2\beta\mu)$. The tensor amplitude up to
leading-order is given by
\begin{equation}
A_{T,q}^{2}=\frac{3q\beta_{q}^{2-2(1-q^{-1})}}{\left(  5\pi\right)  ^{2}}%
\frac{H^{2+2(1-q^{-1})}}{2\zeta_{q}},\label{Ag}
\end{equation}
with $\zeta_{q=1}=\zeta_{q=\frac{2}{3}}=1$\ and \ $\zeta_{q=2}=\frac{2}{3}%
$ \cite{Kim:2004gs}. Finally, the $q$-tensor-scalar ratio from
Eqs.(\ref{Ps}) and (\ref{Ag}) becomes
\begin{equation}
r_q=16\frac{A_{T,q}^{2}}{A_{S,q}^{2}}=16\frac{\epsilon_{1}^{q}}{\zeta_{q}%
}=\frac{16}{\zeta_{q}}\frac{qA^{2\alpha_{0}\alpha_{3}}\alpha_{4}^{1-q}}%
{6\beta_{q}^{2}}\left[  \frac{2\left(  1-f\right)  }{q\left(
\alpha _{1}+1\right)  }\right]  ^{2}\phi^{-2\gamma},\label{r}
\end{equation}
in the patch cosmological models.

From Eqs.(\ref{ns}) and (\ref{r}) we can write the relation
between the tensor-to-scalar
ratio $r_{q}\ $and the spectral index \ $n_{s}^{q}$\ as%
\begin{equation}
r_{q}(n_{s}^{q})=\frac{8(1-f)}{\zeta_{q}(\alpha_{1}+1)}\frac{(1-n_{s}^{q}%
)}{\left[  \frac{(1-f)}{\alpha_{1}+1}\left(  3-\frac{2}{q}\right)
-1\right] }.\label{rq}
\end{equation}

Also, we can write the relation between the number of e-folds  $N$
and the tensor-to-scalar ratio $r_{q}$, from Eqs.(\ref{N}) and
(\ref{r}) as

\begin{equation}
N=\alpha_{8}\left(  \frac{16q\alpha_{4}^{1-q}}{6\zeta_{q}\beta_{q}%
^{2}}\left[  \frac{2(1-f)}{q(\alpha_{1}+1)}\right]
^{2}\;\frac{1}{r_q}-\alpha _{6}\right).\label{NN}
\end{equation}

\begin{figure}[th]
\includegraphics[width=5.0in,angle=0,clip=true]{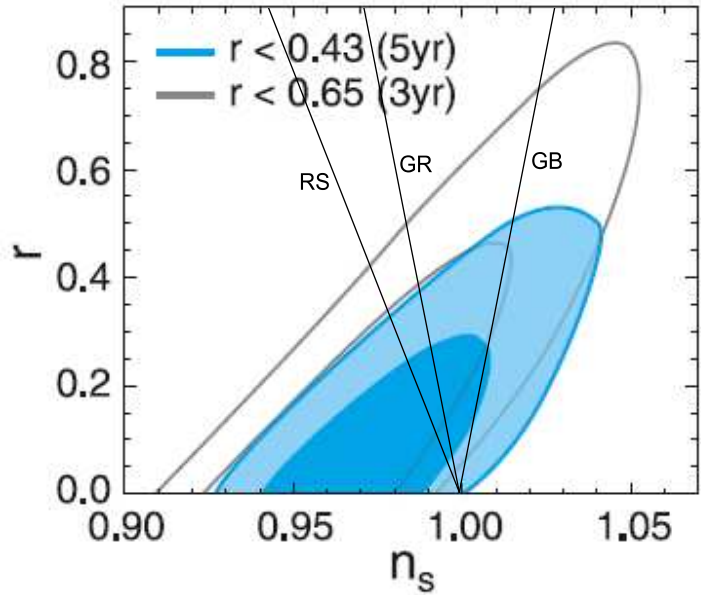}
\caption{ The plot shows $r$ versus $n_s$. Here, we have fixed the
value $f=3/5$. The five-year WMAP data places stronger limits on
the tensor-scalar ratio (shown in blue) than three-year data
(grey) \cite{WMAP3}. The choices  $q = 1, 2, 2/3$, corresponds to
the General Relativity (GR), Randall Sundrum (RS) and Gauss-Bonnet
(GB) regimes, respectively. \label{cou}}
\end{figure}

In Fig.(\ref{cou}) we show the dependence of the tensor-scalar
ratio on the spectral index, from Eq.(\ref{rq}). From left to
right $q$=2 to corresponds (RS), 1 (GR) and 2/3 (GB),
respectively. From Ref.\cite{WMAP3}, two-dimensional marginalized
 constraints (68$\%$ and 95$\%$ confidence levels) on inflationary parameters
$r$, the tensor-scalar ratio, and $n_s$, the spectral index of
fluctuations, defined at $k_0$ = 0.002 Mpc$^{-1}$. The five-year
WMAP data places stronger limits on $r$ (shown in blue) than
three-year data (grey)\cite{Spergel}. In order to write down
values that relate $n_s$ and $r$, we used Eq.(\ref{rq}).  Also we
have used  the value $f=3/5$.  From Eq.(\ref{NN}) and the line of
RS for $q=2$, we observed  that for $f = \frac{3}{5}$, the curve
$r = r(n_s)$ (see Fig. (\ref{cou})) for WMAP 5-years enters the
95$\%$ confidence region for (RS) where the ratio $r_2\simeq
0.33$, which corresponds to the number of e-folds, $N \simeq
47.2$. For $q=1$ (GR), $r_1 \simeq 0.38$ corresponds to $N \simeq
27.3$. For $q=2/3$ (GB), $r_{2/3} \simeq 0.52$ corresponds to $N
\simeq 20.3$. From 68$\%$ confidence region for $q=2$ (RS),
$r_2\simeq 0.28$, which corresponds to $N\simeq$ 57.3. For $q=1$
(GR), $r_1 \simeq 0.25$ corresponds to $N \simeq 41.7$, and for
$q=2/3$ (GB), $r_{2/3} \simeq 0.27$ corresponds to $N \simeq 39$.

 From Eqs.(\ref{P}) and (\ref{At}) we can write the general
relation in GB brane world for the tensor-to-scalar ratio $r_{q}\
$ given by
\begin{equation}
r_q=16\frac{A_{T,q}^{2}}{A_{S,q}^{2}}=\left[\frac{400\,\beta_q^2\,(f-1)^2\,A^{2\alpha_3(q-1)/q}\,\alpha_4^{(q-1)}}{3\,q^2\,(\alpha_1+1)^2\,\beta_1^4}\right]\,\label{ge}
\phi_*^{-2\,\gamma}\,\frac{F_\beta^2(x_*)}{G_\beta^2(x_*)},
\end{equation}
where $x_*\equiv H_*/\mu$.

\begin{figure}[th]
\includegraphics[width=5.0in,angle=0,clip=true]{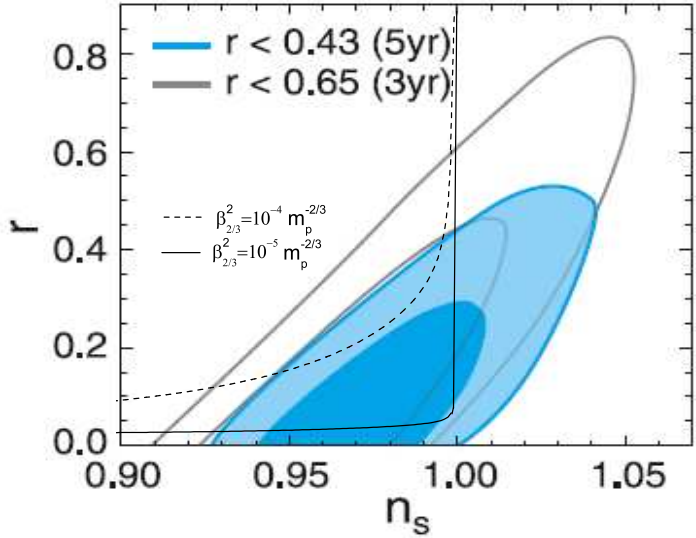}
\caption{ The plot shows $r$ versus $n_s$, for two different
values of the GB parameter $\beta_{2/3}^2$. Here, we have fixed
the values $f=3/5$, $A=10^{-3}\,m_p^{2/3}$ and $\beta=10^{-3}$,
respectively. \label{cou2}}
\end{figure}

In Fig.(\ref{cou2}) we show the dependence of the tensor-scalar
ratio on the spectral index, from Eqs.(\ref{ns}) and (\ref{ge}).
Here, we have taken two different values of the GB parameter
$\beta_{2/3}^2$. In doing this, we have used values $f=3/5$,
$A=10^{-3}\,m_p^{2/3}$, and $\beta=10^{-3}$, respectively. Note
that the Fig.(\ref{cou2}), becomes different to the
Fig.(\ref{cou}) for the case $q=2/3$, when we have used  the
corrections given by Eq.(\ref{ge}) .

Numerically from Eq.(\ref{N}), we observed that for the parameter
$\beta_{2/3}^2=10^{-4} m_p^{-2/3}$ the curve $r = r(n_s)$ (see
Fig. (\ref{cou2})) for WMAP 5-years enters the 95$\%$ confidence
region the ratio $r_{2/3}\simeq 0.41$, which corresponds to the
number of e-folds, $N \simeq 26 $. For $\beta_{2/3}^2=10^{-5}
m_p^{-2/3}$, $r_{2/3} \simeq 0.42$ corresponds to $N \simeq 25$.
Note also that the curve-value $\beta_{2/3}=10^{-4} m_p^{-2/3}$
does not agree with the one-dimensional marginalized
 constraint 68$\%$  confidence level on inflationary parameters
$r$, this is due to its curve is obtained for a given values of
$A$, $\beta$ and $f$.

\section{Conclusion and Final Remarks}

In this work we have studied  an intermediate inflationary
universe model in which the gravitational effects are described by
the Gauss-Bonnet Brane World Cosmology. We study this model by
using the scheme of patch cosmology. In this approach the dynamics
of the scale factor is governed by a modified Friedmann equation
given by $H^2=\beta_q^2\,\rho^q$, where $q=1$ represent GR theory,
$q=2$ describes high energy limit of brane world cosmology, and
$q=2/3$ corresponds to brane world cosmology with a Gauss-Bonnet
correction in the bulk.  We have described different cosmological
models where the matter content is given by a single scalar field
in presence of the  power-law potential. By using  the scalar
potential (see Eq.(\ref{pot})) and from the WMAP five year data,
we have found  constraints on the parameter $A$  for a given
values of $\beta_q$  and $f$ (see Eq.(\ref{A})). In particular,
for $f=2/3$ we obtained that in GR
 ($q=1$) we get $A\simeq 0.0019m_p^{2/3}$. In RS ($q=2$) for the value of
 $\beta_{2}^2=10^{-13}m_p^{-6}$, we have $A\simeq 0.0014
 m_p^{2/3}$, and in the  GB regime ($q=2/3$)  for
 $\beta_{2/3}^2=10^{-3}m_p^{-2/3}$, we get $A\simeq 0.0046
 m_p^{2/3}$. Here we have taken $N=60$ and $\mathcal{P}_{\mathcal{R}}\simeq 2.4\times
 10^{-9}$. In order to bring some explicit results we have taken the
constraint $r-n_s$ plane to first-order in the slow roll
approximation. We noted that the parameter $f$, which lies in the
range $1>f>0$, the model is well supported by the data as could be
seen from Fig.(\ref{cou}).

In this paper, we have not addressed the phenomena of reheating
and possible transition to the standard cosmology  (see e.g.,
Refs.\cite{SRc,u,yo}). A possible calculation for the reheating
temperature in the high-energy scenario would give new constrains
on the parameters of the model.  We hope to return to this point
in the near future.

\begin{acknowledgments}

R.H. was supported by COMISION NACIONAL DE CIENCIAS Y TECNOLOGIA
through FONDECYT Grant N$^{0}$.   1090613,  also from  PUCV
DI-PUCV 2009.
\end{acknowledgments}


\begin{thebibliography}{99}

\bibitem{R1}A. Guth , Phys. Rev. D {\bf23}, 347 (1981); A.A. Starobinsky, Phys. Lett.
     B {\bf91}, 99 (1980); A.D. Linde,  Phys. Lett. B {\bf108}, 389 (1982); {\it idem} Phys. Lett.
    B {\bf129}, 177 (1983); A. Albrecht  and P. J. Steinhardt, Phys. Rev. Lett.
   {\bf48},1220 (1982); K. Sato,  Mon. Not. Roy. Astron. Soc. {\bf195}, 467 (1981).



\bibitem{R2}V.F. Mukhanov  and G.V. Chibisov , JETP Letters {\bf33}, 532(1981); S. W. Hawking,
     Phys. Lett. B {\bf115}, 295 (1982); A. Guth  and S.-Y. Pi,
    Phys. Rev. Lett. {\bf49}, 1110 (1982); A. A. Starobinsky,  Phys. Lett. B {\bf117},
    175 (1982); J.M. Bardeen,  P.J. Steinhardt  and M.S. Turner,  Phys.
    Rev.D {\bf28}, 679 (1983).


\bibitem{R4}G. Smoot,  et al. Astrophys. J. Lett. {\bf396}, L1 (1992).







\bibitem{IC} A. Guth, Phys. Rev. D {\bf 23}, 347 (1981);
A. Albrecht and P. J. Steinhardt, Phys. Rev. Lett. {\bf 48}, 1220
(1982). ; A. D. Linde, Phys. Lett. {\bf 108B}, 389 (1982), Phys.
Lett. {\bf 129B}, 177 (1983).

\bibitem{libro} A. D. Linde, \emph{Particle Physics and Inflationary
Cosmology}. Harwoord, Chur, Switzerland, (1990).
arXiv:hep-th/0503203.


\bibitem{Barrow1} J. D Barrow,
Phys. Lett. B {\bf 235}, 40 (1990); J. D Barrow and P. Saich,
Phys. Lett. B {\bf 249}, 406 (1990);A. Muslimov, Class. Quantum
Grav. {\bf 7}, 231 (1990); J. D Barrow and A. R. Liddle, Phys.
Rev. D {\bf 47}, R5219 (1993); A. D. Rendall, Class. Quantum Grav.
{\bf 22}, 1655 (2005).

\bibitem{yo3} S.~del Campo, R.~Herrera and A.~Toloza,
  Phys.\ Rev.\  D {\bf 79}, 083507 (2009)


\bibitem{yo4} S.~del Campo and R.~Herrera,
  JCAP {\bf 0904}, 005 (2009)


\bibitem{yo5}S.~del Campo and R.~Herrera,
  Phys.\ Lett.\  B {\bf 670}, 266 (2009)


\bibitem{1}J.E.Lidsey, astro-ph/0305528; J.E.Lidsey, D.Wands and
E.J.Copeland, Phys. Rep. \textbf{337}, 343 (2000); M.Gasperini and
G.Veneziano, Phys. Rep. \textbf{373}, 1 (2003); M.Quevedo, Class.
Quant. Grav. \textbf{19}, 5721 (2002). 



\bibitem {17} D. G. Boulware and S. Deser, Phys.Rev. Lett. {\bf55}, 2656
(1985); Phys. Lett. B {\bf175}, 409 1986).

\bibitem{18}G. Gognola, E. Eizalde, S. Nojiri, S. D. Odintsov
and S. Zerbini, Phys. Rev. D {\bf73}, 084007 (2006).

\bibitem{19} I. Antoniadis, J. Rizos and K. Tamvakis, Nucl.Phys. B {\bf415},
497 (1994).

\bibitem{20} S. Mignemi and N. R. Steward, Phys. Rev. D {\bf47}, 5259 (1993);
P. Kanti, N. E. Mavromatos, J. Rizos, K. Tamvakis and E.
Winstanley, Phys. Rev. D {\bf54}, 5049 (1996); Ch.-M Chen, D. V.
Gal'tsov and D. G. Orlov, Phys. Rev. D {\bf75}, 084030 (2007).

\bibitem{21} S. Nojiri, S. D. Odintsov and M. Sasaki, Phys. Rev. D {\bf71},
123509 (2004).



\bibitem{Meng:2003pn}  X.~H.~Meng and P.~Wang,
  Class.\ Quant.\ Grav.\  {\bf 21}, 2527 (2004)
 \bibitem{Lidsey:2003sj}
  J.~E.~Lidsey and N.~J.~Nunes,
  Phys.\ Rev.\  D {\bf 67}, 103510 (2003).



\bibitem{Calcagni:2004as}G.~Calcagni and S.~Tsujikawa,
Phys.\ Rev.\  D {\bf 70}, 103514 (2004).




\bibitem{Kim:2004gs}
  H.~Kim, K.~H.~Kim, H.~W.~Lee and Y.~S.~Myung,
  Phys.\ Lett.\  B {\bf 608}, 1 (2005).
\bibitem{Kim:2004jc}
  K.~H.~Kim and Y.~S.~Myung,
  JCAP {\bf 0412}, 004 (2004).
\bibitem{Kim:2004hu}
  K.~H.~Kim and Y.~S.~Myung,
  Int.\ J.\ Mod.\ Phys.\  D {\bf 14}, 1813 (2005).
\bibitem{Murray:2006fw}
  B.~M.~Murray and Y.~S.~Myung,
  Phys.\ Lett.\  B {\bf 642}, 426 (2006).
\bibitem{22} A. K. Sanyal, Phys. Lett. B {\bf 645},1 (2007).



\bibitem{2} N.Arkani-Hamed, S.Dimopoulos and G.Dvali, Phys. Lett. B
\textbf{429}, 263 (1998); I.Antoniadis, N.Arkani-Hamed,
S.Dimopoulos and G.Dvali, Phys. Lett. B \textbf{436}, 257 (1998);
L.Randall and R.Sundrum, Phys. Rev. Lett. \textbf{83}, 3370
(1999).


\bibitem{witten}  P. Horava and E. Witten, Nucl.Phys.B {\bf 475}, 94 (1996).

\bibitem{witten2}P. Horava and E. Witten, Nucl.Phys.B {\bf 460}, 506 (1996).


\bibitem{lecturer}  J. Lidsey, Lect.\ Notes Phys.\  {\bf 646}, 357
(2004).

\bibitem{lecturer2} P. Brax, C. van de Bruck. Class.Quant.Grav.{\bf 20}, R201-R232
(2003).
\bibitem{lecturer3} E. Papantonopoulos, Lect.Notes Phys. {\bf 592}, 458
(2002).


\bibitem{Ch} C.Charmousis and J-F. Dufaux, Class.\ Quant.\ Grav.\  {\bf 19}, 4671
(2002); J.E. Lidsey and N.J. Nunes, Phys. Rev. D, \textbf{67},
103510 (2003); Kei-ichi Maeda, Takashi Torii, Phys.\ Rev.\  D {\bf
69}, 024002 (2004); J.-F. Dufaux, J. Lidsey, R. Maartens, M. Sami,
Phys.\ Rev.\  D {\bf 70}, 083525 (2004);
%
B. Abdesselam and N. Mohammedi, Phys. Rev. D \textbf{65}, 084018
(2002).


\bibitem{Davis:2002gn}
  S.~C.~Davis,
  Phys.\ Rev.\  D {\bf 67}, 024030 (2003).


\bibitem{Gravanis:2002wy}
  E.~Gravanis and S.~Willison,
  Phys.\ Lett.\  B {\bf 562}, 118 (2003).


\bibitem{Calcagni:2004bh}
  G.~Calcagni,
  Phys.\ Rev.\  D {\bf 69}, 103508 (2004).

\bibitem{yo1} S.~del Campo, R.~Herrera, J.~Saavedra and P.~Labrana,
  Annals Phys.\  {\bf 324}, 1823 (2009).


\bibitem{yo2} S.~del Campo, R.~Herrera, J.~Saavedra, P.~Labrana and C.~Leiva,
Mod. Phys. Lett. A {\bf 24}, 2445 (2009).

\bibitem{Od}J.~E.~Lidsey, S.~Nojiri and S.~D.~Odintsov,
  JHEP {\bf 0206}, 026 (2002).


\bibitem{Od2}S.~Nojiri, S.~D.~Odintsov and S.~Ogushi,
  Int.\ J.\ Mod.\ Phys.\  A {\bf 17}, 4809 (2002).


\bibitem{Od3}S.~Nojiri, S.~D.~Odintsov and S.~Ogushi,
  Phys.\ Rev.\  D {\bf 65}, 023521 (2002).


\bibitem{pp} J.~F.~Dufaux, J.~E.~Lidsey, R.~Maartens and M.~Sami,
  Phys.\ Rev.\  D {\bf 70}, 083525 (2004).

\bibitem{Chung:1999zs}
  D.~J.~H.~Chung and K.~Freese,
  Phys.\ Rev.\  D {\bf 61}, 023511 (2000).

\bibitem{13}B. Ratra and P. J. E. Peebles, Phys. Rev. D {\bf 37}, 3406
(1988); I. Zlatev, L. Wang, and P. J. Steinhardt, Phys. Rev. Lett.
{\bf 82}, 896 (1999).

\bibitem{14}J.D. Barrow and K.I. Maeda, Nucl. Phys. B {\bf 341}, 294
(1990).


\bibitem{Maartens:1999hf}
  R.~Maartens, D.~Wands, B.~A.~Bassett and I.~Heard,
  Phys.\ Rev.\  D {\bf 62}, 041301 (2000).


%

%




\bibitem{WMAP3} J.~Dunkley {\it et al.}  [WMAP Collaboration],
  Astrophys.\ J.\ Suppl.\  {\bf 180}, 306 (2009);  G.~Hinshaw {\it et al.},
  Astrophys.\ J.\ Suppl.\  {\bf 180}, 225 (2009).


\bibitem{Spergel} D.~N.~Spergel {\it et al.}, Astrophys.\ J.\
Suppl.\  {\bf 170}, 377 (2007).

















\bibitem{SRc}  S.~del Campo and R.~Herrera,
Phys.\ Rev.\  D {\bf 76}, 103503 (2007).


\bibitem{u}  E.~J.~Copeland, A.~R.~Liddle and J.~E.~Lidsey,
  Phys.\ Rev.\  D {\bf 64}, 023509 (2001);
  E.~J.~Copeland and O.~Seto,
  Phys.\ Rev.\  D {\bf 72}, 023506 (2005).


\bibitem{yo} C.~Campuzano, S.~del Campo and R.~Herrera,
  JCAP {\bf 0606}, 017 (2006); C.~Campuzano, S.~del Campo and R.~Herrera,
  Phys.\ Lett.\  B {\bf 633}, 149 (2006);  C.~Campuzano, S.~del Campo and R.~Herrera,
  Phys.\ Rev.\  D {\bf 72}, 083515 (2005)
  [Erratum-ibid.\  D {\bf 72}, 109902 (2005)].





\end{thebibliography}
\end{document}